\documentclass{article}

\PassOptionsToPackage{numbers, compress}{natbib}

\usepackage[preprint]{neurips_2024}

\usepackage[utf8]{inputenc} 
\usepackage[T1]{fontenc}    
\usepackage{hyperref}       
\usepackage{url}            
\usepackage{booktabs}       
\usepackage{amsfonts}       
\usepackage{nicefrac}       
\usepackage{microtype}      
\usepackage[table]{xcolor} 
\usepackage{booktabs} 
\usepackage{colortbl} 
\usepackage{graphicx}
\usepackage{algorithm}
\usepackage{algpseudocode}
\usepackage{tabularray}
\usepackage{amsmath}
\usepackage{amssymb}
\usepackage{mathtools}
\usepackage{subfigure}
\usepackage{amsthm}
\usepackage{natbib}
\usepackage[capitalize,noabbrev]{cleveref}

\usepackage[table]{xcolor} 
\usepackage{booktabs} 
\usepackage{colortbl} 

\definecolor{LightGray}{gray}{0.9} 
\usepackage{nameref,hyperref}

\title{NERULA: A Dual-Pathway Self-Supervised Learning Framework for Electrocardiogram Signal Analysis}

\author{%
  Gouthamaan Manimaran\\
  Department of Health Technology\\
  Technical University of Denmark\\
  Copenhagen, Denmark \\
  \texttt{gouma@dtu.dk} \\
   \And
  Sadasivan Puthusserypady\\
  Department of Health Technology\\
  Technical University of Denmark\\
  Copenhagen, Denmark \\
  \texttt{sapu@dtu.dk} \\
  \And
  Helena Domínguez\\
  Department of Cardiology\\
  Bisperbjerg Hospital\\
  Copenhagen, Denmark \\
  \texttt{maria.helena.dominguez.vall-lamora.02@regionh.dk} \\
  \And
  Adrian Atienza\\
  Department of Health Technology\\
  Technical University of Denmark\\
  Copenhagen, Denmark \\
  \texttt{adar@dtu.dk} \\
  \And
  Jakob E. Bardram\\
  Department of Health Technology\\
  Technical University of Denmark\\
  Copenhagen, Denmark \\
  \texttt{jakba@dtu.dk} \\
}

\begin{document}

\maketitle

\begin{abstract}

  Electrocardiogram (ECG) signals are critical for diagnosing heart conditions and capturing detailed cardiac patterns. As wearable single-lead ECG devices become more common, efficient analysis methods are essential. We present NERULA (Non-contrastive ECG and Reconstruction Unsupervised Learning Algorithm), a self-supervised framework designed for single-lead ECG signals. NERULA's dual-pathway architecture combines ECG reconstruction and non-contrastive learning to extract detailed cardiac features. Our 50\% masking strategy, using both masked and inverse-masked signals, enhances model robustness against real-world incomplete or corrupted data. The non-contrastive pathway aligns representations of masked and inverse-masked signals, while the reconstruction pathway comprehends and reconstructs missing features. We show that combining generative and discriminative paths into the training spectrum leads to better results by outperforming state-of-the-art self-supervised learning benchmarks in various tasks, demonstrating superior performance in ECG analysis, including arrhythmia classification, gender classification, age regression, and human activity recognition. NERULA's dual-pathway design offers a robust, efficient solution for comprehensive ECG signal interpretation.

\end{abstract}

\section{Introduction}
\label{intro}

In the realm of medical signal processing, the Electrocardiogram (ECG) stands as a cornerstone for cardiac health assessment. The ECG's capacity to non-invasively capture the heart's electrical activity renders it an indispensable tool in diagnosing a wide array of cardiac conditions. However, the nuanced interpretation of ECG signals, traditionally a domain of specialized clinicians, presents a significant challenge. The advent of machine learning, particularly in the field of deep learning, has opened new avenues for automated ECG analysis, offering the potential to augment clinical expertise with data-driven insights.

Recent advances in machine learning have predominantly leaned on supervised learning paradigms, where models are trained on large datasets with predefined labels. Despite their success, these methods are constrained by the availability of high-quality, annotated datasets, which are expensive and time-consuming to produce, especially in the medical domain. Furthermore, supervised models often lack the ability to generalize to new, unseen data, a limitation particularly pronounced in healthcare applications where patient variability is high.

Enter self-supervised learning (SSL), a paradigm that learns representations from data without explicit labels. In the context of ECG analysis, SSL presents a promising avenue to circumvent the limitations of supervised approaches. By leveraging the inherent structure within the ECG data, SSL methods can unearth robust and generalizable features, paving the way for models that are both data-efficient and adaptable to diverse patient populations.

Our work introduces a novel approach to self-supervised representation learning for ECG signals. At its core, our method utilizes a dual-pathway architecture comprising an ECG reconstruction path and a non-contrastive learning path, both originating from a shared encoder. This design is engineered to capture the intricate patterns within ECG signals, facilitating a comprehensive understanding of cardiac rhythms and potential anomalies.

A key innovation in our approach is the handling of input signals. By introducing a 50\% masking strategy, along with the use of the inverse of the masked signal, our model is tasked with reconciling these two complementary views of the same data. This strategy not only enhances the robustness of the learned representations but also aligns closely with real-world scenarios where incomplete or corrupted ECG data is commonplace.

The non-contrastive learning path of our model is designed to ensure that the representations of the masked and inverse-masked signals converge, thus learning consistent features across different transformations of the data. Simultaneously, the ECG reconstruction path, guided by Huber loss, assists the network in understanding and reconstructing the missing features of the ECG signal, an aspect critical for comprehensive signal interpretation.

The contributions of our work are as follows:
\begin{itemize}
    \item We introduce \textit{NERULA}, a novel method in representation learning for Time-Series signals that merge a Generative (Signal Reconstruction) and Discriminative (Non-Contrastive Learning) pathway that leads to more efficient learning of latent features.
    \item In non-contrastive learning of ECGs, we emphasize the importance of what is chosen as \textit{positive} pairs by showing improved results in random masking of signals over other signal augmentation types like in BYOL or adjacent and non-overlapping segments like in CLOCS.
    \item We show that NERULA outperforms other state-of-the-art SSL methods such as BYOL, SimCLR, and CLOCS in downstream linear evaluation tasks which include arrhythmia classification, gender classification, age regression, and human activity recognition benchmarks.
\end{itemize}

\section{Related Work}
In this section, we introduce some notable works in self-supervised learning as a whole and also within the context of time-series/ECG signal processing. 

\textbf{Contrastive and Non-Contrastive Learning}: There have been many recent works on self-supervised learning that leverage the use of this concept of `positive' and/or `negative' pairs to learn meaningful features from the input. One of the most notable works of contrastive learning in computer vision is \textit{SimCLR} \cite{simclr}, which uses an augmented view of the input as \textit{positive} and a different image as \textit{negative}. One of the drawbacks of this method is that the choice of these negative pairs is random across the whole dataset, and some instances, even if they are from the same class (e.g. bird, tree, etc) might be considered as negative thereby hindering the learning curves. To overcome this, \textit{BYOL} \cite{byol} make use of only the positive pairs and use an additional network (prediction network) which is a moving average of the main network (target network) to align these two predictions using a non-contrastive loss function.  \\

\begin{figure}[!ht]
\centering
\subfigure[Original Signal]{
  \includegraphics[width=0.45\columnwidth]{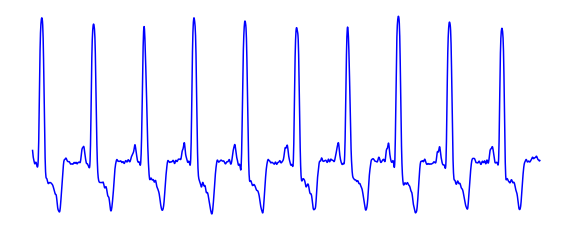}
}
\subfigure[Masked Signal]{
  \includegraphics[width=0.45\columnwidth]{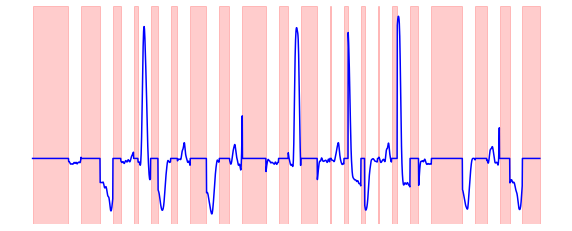}
}
\subfigure[Signal masked in random positions]{
  \includegraphics[width=0.45\columnwidth]{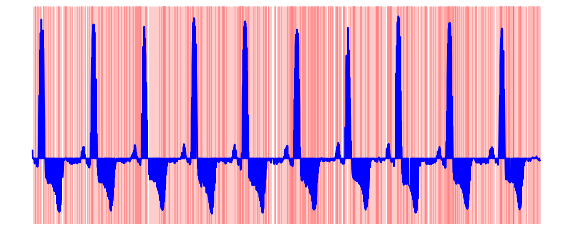}
}
\subfigure[Inverted-Masked Signal]{
  \includegraphics[width=0.45\columnwidth]{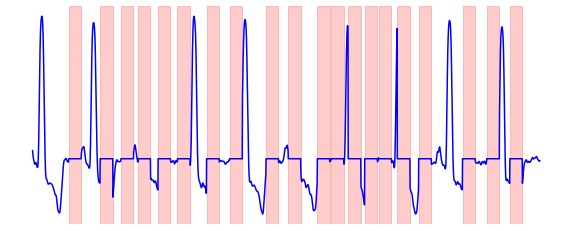}
}

\caption{Signal Masking Strategies for Non-Contrastive Learning}
\label{masking}
\end{figure}

\textbf{Reconstruction Learning}: Meaningful latent representations can also be learned by methods that are able to reconstruct missing parts of the input signal. While this train of thought does wonders in the field of natural language processing in methods like BERT \cite{bert} and ERNIE \cite{ernie} that try to predict missing/masked words in a sentence, initially there had been no impressive work in computer vision or time-series processing that show this is a better approach than the widely used contrastive learning. Some recent work in computer vision shows that this approach is worth exploring and is mainly using Masked Autoencoder (MAE) approaches like the work by \textit{He et al.} \cite{mae} in (2 dimensional) images and \textit{Feichtenhofer et al.} \cite{mae_st} in (3 dimensional) videos. Following these methods, there has also been a similar implementation of the \cite{mae} method in ECG processing by \textit{Hu et al.} \cite{mae_ecg}, where they pass only the non-masked patches of the ECG signal into the encoder and then pass the latent representations of this along with the masked tokens to reconstruct the signal. They show that this approach was able to get higher accuracies than \textit{BYOL} and \textit{SimCLR} in signal classification benchmarks. A more generalized time-series approach of the same method has been done by \textit{Li et al.} \cite{timae} which we use for comparisons in this paper.  \\

\textbf{Addressing the Gap}: Both the methods mentioned above have been used extensively in computer vision, and some methods have also been able to incorporate learning from both these methods into one hybrid learning curve like \cite{internvideo} that have loss functions flowing from two separate heads (contrastive and reconstruction) into a single encoder. However, there has been little to no work in combining the generative path (reconstruction) and discriminative path (contrastive) for time-series modeling.
It is also worth noting that one cannot reconstruct a time-series signal the same way as an image is done. Most works split the images into non-overlapping patches with discrete but learnable positional encoding to feed into the transformer. While this may make sense for images, time-series signals are inherently different, not to mention a whole missing dimension (if leads are not involved in ECGs). This can be noticed in many classification tasks like the PhysioNet 2021 challenge \cite{physio2021}, where most of the winning solutions are convolution-based algorithms and not transformers - this is mainly due to the continuous temporal nature of these signals which are not suitable to be split into patches as a preprocessing step for transformer architecture. 
Another significant gap is the choice of data augmentation of a time-series signal. Augmentations used in computer vision like the Random Crop \cite{random_crop} or Inverting an axis change the inherent nature of the signals unlike in computer vision where they still make sense. We address such gaps not only between self-supervised methods but also the inherent difference between the domains.

\section{Background}

\textbf{Self-Supervised Learning} (SSL) is a form of unsupervised learning where the data itself provides the supervision. In SSL, the model learns to predict any part of the data from any other part of it. The general objective function can be represented as:
\begin{equation}
    L(\theta) = \sum_{x \in \mathcal{D}} \ell(f_\theta(x), x),
\end{equation}
where \( L(\theta) \) is the loss function for a model parameterized by \( \theta \), \( \mathcal{D} \) represents the dataset, \( f_\theta \) is the representation function, and \( \ell \) is a suitable loss measuring the prediction quality.

\textbf{Contrastive vs Non-Contrastive Learning}: In contrastive learning, models learn to distinguish between similar (positive) and dissimilar (negative) pairs of data points. Non-contrastive learning, on the other hand, does not explicitly use negative pairs and focuses on learning representations that bring similar samples closer in the embedding space. The contrastive and non-contrastive loss functions can be described as follows:

Contrastive Loss:
\begin{equation}
    L_{ce} = -\log \frac{\exp(sim(z_i, z_j) / \tau)}{\sum_{k=1}^{N} \mathbb{1}_{[k \neq i]} \exp(sim(z_i, z_k) / \tau)}
\end{equation}

Non-Contrastive Loss:
\begin{equation}
    L_{nce} = \mathbb{E}_{x, x^+}\left[ d(f_\theta(x), f_\theta(x^+)) \right]
\end{equation}

where \( sim(\cdot) \) is a similarity measure, \( \tau \) is a temperature scaling parameter, \( z_i \), \( z_j \), and \( z_k \) are different data representations, and \( d(\cdot) \) is a distance metric.

Non-contrastive learning is crucial in scenarios where constructing negative pairs is challenging or may introduce biases and we argue that this is the case for physiological time-series signals like Electrocardiograms (ECG). It is also computationally more efficient, avoiding the need for large batches or memory banks typically used in contrastive learning.

\textbf{Signal Reconstruction} is an emerging paradigm in Self-Supervised Learning (SSL) where the primary task is to reconstruct the input signal from its corrupted or partial version. This approach is grounded in the assumption that a model capable of accurate reconstruction has effectively learned meaningful and robust representations of the data.

In signal reconstruction, an input signal \( x \) is first transformed into a corrupted or partial version \( \tilde{x} \), often through masking or corruption function \( C \). The model, typically an encoder-decoder architecture, then attempts to reconstruct the original signal \( x \) from \( \tilde{x} \). The objective function for signal reconstruction in SSL can be formulated as:

\begin{equation}
    L_{recon}(\theta) = \mathbb{E}_{x \in \mathcal{D}, \tilde{x} = C(x)} \left[ \| f_\theta(\tilde{x}) - x \|^2 \right]
\end{equation}

where \( L_{recon}(\theta) \) is the reconstruction loss, \( f_\theta \) is the reconstruction function parameterized by \( \theta \), and \( \| \cdot \|^2 \) denotes the squared error between the reconstructed and original signals.

This approach is particularly effective in domains where the underlying structure of the data is complex yet consistent, such as in speech, audio processing, and time-series analysis. The key advantage of signal reconstruction as an SSL method lies in its ability to learn intricate data patterns without the need for explicit labels, thereby enabling models to generalize better to unseen data.

\section{Method}

\subsection{Temporal Input Masking}

\begin{figure}[!ht]
\centering

\includegraphics[width=0.99\columnwidth]{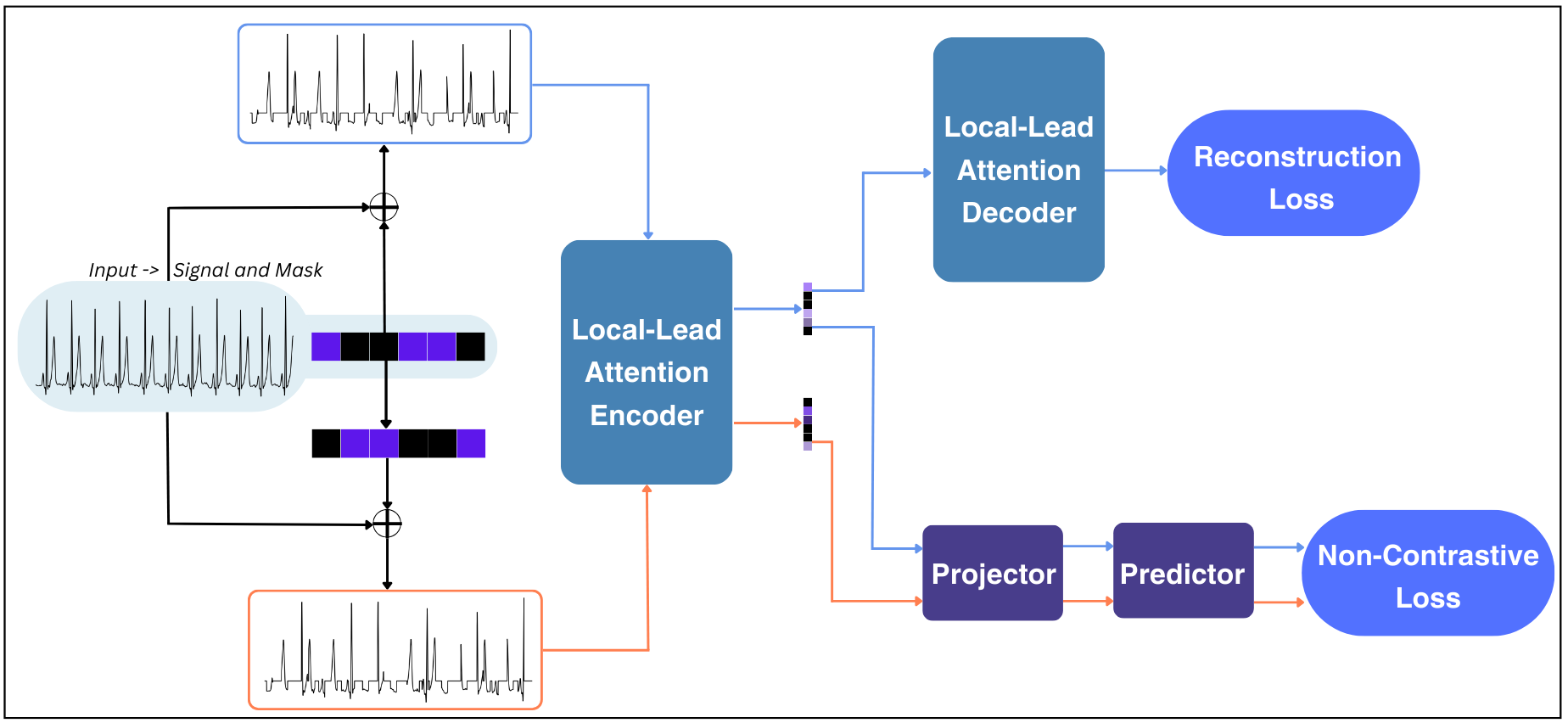}%

\caption{\textbf{NERULA Structure Overview}. The input to the algorithm is the raw signal and a corresponding random binary mask. From these, we generate a pair of signals with complimentary masks. Our encoder generates latent representations of both these signals after which they are passed through a non-contrastive setup and one of the signals is reconstructed with a decoder.}
\label{method}
\end{figure}

In this method, we mask exactly 50\% of the input for both the tasks - non-contrastive and reconstruction. In the latter task, the masked input is passed through a shared encoder and is reconstructed to its original input. For non-contrastive learning, the two positive pairs are both the same signal but masked at complementary positions as shown in Fig \ref{masking} (b) and (c). The masking strategy used here is of utmost importance. Randomly masking any 50\% positions of the signal will not contribute much to both tasks. As shown in Fig \ref{masking} (d), we can see that even the randomly masked signal still looks like the original signal, and can be reconstructed to a reasonable degree by just using linear interpolation and can be greatly improved in precision if using kernels of greater sizes. Thus, this is not the ideal strategy to mask signals. Instead, we mask \textit{patches} of a signal. The position of the patches, The length, and the number of these patches are random, and we set the number of such patches to be between 15-30. This was chosen as having patches that are too large, hinders the reconstruction of the signal and it fails in such areas.  

\begin{table}[!ht]
\begin{center}
\caption{Difference in performance based on the type of \textit{positive} pairs used for non-contrastive learning. The metrics reported in this table are F1 scores.}
\begin{tabular}{lcc} 
\toprule
Augmentation used & SVC & Random Forest   \\ 
\midrule
BYOL & 0.45 & 0.30  \\
CLOCS & 0.44 & 0.32   \\
\rowcolor{LightGray}
NERULA & \textbf{0.49} & \textbf{0.42}  \\
\bottomrule
\end{tabular}
\label{tab:ablation_augmentation}
\end{center}
\end{table}

In Table \ref{tab:ablation_augmentation}, we show the effect of using different types of augmentation techniques to generate positive pairs for non-contrastive learning in ECGs. In all these experiments, we use ResNet \cite{resnet} as the encoder and a non-contrastive loss function. The following are the augmentation techniques for each method:
\begin{enumerate}
    \item \textbf{BYOL} uses standard data augmentation techniques like flipping across the x or y axis, random crop, and gaussian noise.
    \item \textbf{CLOCS} uses two non-overlapping and adjacent segments from an ECG recording.
    \item \textbf{NERULA} uses a temporal random mask where the second signal is generated from the inverted mask.
\end{enumerate}

Through this ablation, we can see the effect of different \textit{augmentation} techniques on classification tasks, and it is clear that for ECG modeling, random masking gives superior performance.

\subsection{Latent Space Masking}
Existing methods in Masked Image Modelling (MIM) like \cite{mae, multiscale, mim2, mim3, mim4} only pass in the visible patches to the encoder and later pass in the masks at the correct positions for the decoding stage. We do not do this as our mask patch sizes are not equal in length and vary in both number and size. Instead, we pass in the whole input as a single signal - both masked and unmasked parts into the encoder. As we want only the learned latent representations along with masked patches at the end, we make use of masked convolutions at every layer of the encoder. As the shape of the latent space reduces with encoder depth, we interpolate the mask size to that of the latent vector and multiply them to get the final masked latent representation. Thus, the relative position of the masks does not change and the resultant latent vectors still carry information about the unmasked regions of the signal. This whole vector is then passed onto the decoder for reconstruction, or non-contrastive loss is applied to these vectors.

\begin{table*}[!bht]
\centering
\caption{Cumulative Improvement in F1 Scores: Gradual Component Addition from BYOL to NERULA}
\begin{tabular}{clccc} 
\toprule
& Method & Accuracy  & F1 score & Change in F1 Score  \\ 
\midrule
 & BYOL & 0.56 & 0.30 & -  \\
1.& +Masking as Augmentation & 0.60 & 0.42 & \textcolor{green}{+12\%}  \\
2.& +Feature Masking with Local-Lead Attention & 0.64 & 0.48 & \textcolor{green}{+6\%} \\
3.& +Reconstruction = NERULA & 0.78 & 0.74 & \textcolor{green}{+26\%} \\
\bottomrule
\label{cumulative_table}
\end{tabular}
\end{table*}

\subsubsection{Choice of Encoder}
To be able to have these requirements for masking latent space, as well as a fast and low computationally demanding encoder, we leverage the work done by \cite{locallead}, which is an extension of the performer \cite{performer} model for time-series or ECG processing. In this architecture, a sliding window is used to compute attention weights instead of doing global processing, thus reducing the time complexity from \(O(T^2D + D^2T)\) to \(O(W^2T D + D^2T)\), where \(T\) is number of time steps, \(D\) is the dimension of features and \(W\) is the local window size (\( \ll T\)).

\subsubsection{Non-Contrastive Modelling}
Due to the masking of latent representations at every layer, we get a final representation vector of size L (where L$=$128) with masks that are complimentary for the two positive pairs. Thus, every position that has a positive value in resultant vector 1, will be zero in resultant vector 2, and vice-versa. We apply the non-contrastive loss to this pair of vectors. 

Consider a data point \( x \) and its two masked versions \( x_i \) and \( x_j \) generated through the masking strategy explained in the above sections where if \( m \) is the mask of the same length as \( x \), then: 
\begin{equation}
\ x_i = m \times x,  
\end{equation}
\begin{equation}
\ x_j = (1-m) \times x.
\end{equation}
Non-contrastive learning aims to learn an embedding function \( f(\cdot) \) such that the representations \( f(x_i) \) and \( f(x_j) \) are close in the embedding space. This can be formalized as minimizing a loss function \( \mathcal{L} \), which can be defined in various ways depending on the specific method. We use the cosine similarity measure to minimize this function:
\begin{equation}
\mathcal{L} = -\mathbb{E}_{x, x_i, x_j}\left[ \frac{f(x_i) \cdot f(x_j)}{\| f(x_i) \| \| f(x_j) \|} \right],
\end{equation}
where \( \mathbb{E} \) denotes the expectation over the data distribution. This formulation encourages the model to learn embeddings that are invariant to the augmentations or masking applied to the data, thus capturing the underlying semantics of the data without relying on contrasting with negative examples. This approach has been shown to be effective in various tasks, such as image and text representation learning, where constructing meaningful negative pairs can be challenging such as in ECG processing.

\subsection{Reconstruction Path}

In the reconstruction of ECG signals, We use the Huber Loss function due to the inherent attributes of ECG signals like noise and outliers, as different segments in ECGs like the QRS peaks, P-wave, etc might be abnormal and may be considered an outlier which hinders efficient learning with Mean-Squared-Error (MSE) loss. The Huber Loss is less sensitive to outliers than the MSE loss because it combines the properties of both the mean squared error and the mean absolute error. It uses a squared error when the error is small and an absolute error when the error is large, which makes it robust to outliers. The Huber Loss for a set of predictions can be formally defined as:

\begin{equation}
L_{\delta}(y, \hat{y}) = \begin{cases} 
\frac{1}{2}(y - \hat{y})^2 & \text{for } |y - \hat{y}| \leq \delta, \\
\delta \cdot (|y - \hat{y}| - \frac{1}{2}\delta) & \text{otherwise}.
\end{cases}
\end{equation}

Here, \( y \) represents the true value of the signal, \( \hat{y} \) is the predicted or reconstructed value, and \( \delta \) is a threshold that determines the transition point between the mean squared error and the mean absolute error. The use of the Huber Loss in ECG signal reconstruction helps in achieving a balance between the sensitivity to large errors and robustness to small errors, making it particularly useful in handling anomalies and preserving the integrity of the cardiac signal for reliable clinical analysis. 

As we use a dual-pathway approach in our method, integrating both non-contrastive learning and signal reconstruction- To balance the contributions of these two learning paradigms, we assign weights to their respective loss functions. Specifically, the loss from the non-contrastive pathway is weighted by 1, and the loss from the reconstruction pathway is assigned a weight of 10. This decision was guided by preliminary observations and theoretical considerations about the relative importance and scale of the two loss functions in our specific model architecture and dataset. It is important to note that these weights were chosen based on initial intuitions and practical considerations, rather than through exhaustive hyperparameter tuning. As such, they represent a starting point for exploration rather than an optimized solution. Future work may involve a more systematic investigation into the optimal balance of these weights, potentially leading to further improvements in the model's performance.

\section{Experiments and Results}
\label{sec5}

\subsection{Datasets}
\label{section:datasets}
\textbf{Training} For all our analysis, we chose to train our and other algorithms on the PhysioNet 2020 \cite{physio2020} dataset which consists of 43,101 12-lead ECG signals from over 6 hospitals in 4 countries and 3 continents. Each of these recordings can be mapped to one or more classes with a total of 27 arrhythmia classes. However since we use this only for self-supervised pre-training, we do not make use of the labels. Also, since this is a 12-lead dataset, we only take the lead-II of every recording, since lead-II is the most similar to the \textit{in-the-wild} single-lead Holter monitors.

\textbf{Testing} To evaluate all algorithms on Cardiac \textbf{Arrhythmia} classes, we chose the PhysioNet 2017 \cite{physio2017} dataset which has a total of three classes. This dataset consists of 8,528 recordings and has a preset split of train and validation, which we use to both fit the machine learning model (SVC, Random Forest, and Logistic Regression) and get the scores(Accuracy, F1, AUC). \\
\textbf{HAR}: To complement our testing of arrhythmia classes, we also test on other time-series signals to show the learning and generalization ability of our method. For this, we use the human activity recognition dataset \cite{uci_har} which contains data from 30 volunteers aged between 19-48, and find the activity class from the actimetry sensor data. This dataset has 3-axis accelerometer and gyroscope data along with 6 target classes like walking, sitting, etc. 

We apply a consistent data split methodology for all our experiments, using a train (pretraining), validation (fitting the ML model), and test (computing final scores) split.

\begin{table*}[!ht]
\centering
\caption{Evaluation on arrhythmia classification tasks on the PhysioNet 2017 benchmark. The accuracy, F1 score, and AUC of a 1D ResNet-50 are provided for context. Bold values reflect the top-performing method among the self-supervised algorithms.}
\label{comparison_table}
\begin{tabular}{lccccc}
\toprule
 & \multicolumn{2}{c}{SVC} & \multicolumn{2}{c}{Random Forest} & Logistic Regression \\
\cmidrule(lr){2-3} \cmidrule(lr){4-5} \cmidrule(l){6-6}
Method & Accuracy & F1 Score & Accuracy & F1 Score & AUC \\
\midrule
ResNet (Supervised) & 0.76 & 0.75 & - & - & 0.89 \\
\midrule
BYOL & 0.52 & 0.45 & 0.56 & 0.30 & 0.68 \\
SimCLR & 0.53 & 0.49 & 0.57 & 0.27 & 0.67 \\
Ti-MAE & 0.51 & 0.50 & 0.56 & 0.25 & 0.66 \\
CLOCS & 0.51 & 0.45 & 0.59 & 0.36 & 0.66 \\
\rowcolor{LightGray}
NERULA & \textbf{0.56} & \textbf{0.50} & \textbf{0.78} & \textbf{0.74} & \textbf{0.79} \\
\bottomrule
\end{tabular}
\end{table*}

\subsection{Evaluation}
\label{eval}
We conduct all our experiments on PyTorch and one NVIDIA GeForce RTX 4090 GPU. The Adam \cite{adam} optimizer was used with a learning rate of \(\mathrm{1e-4}\) for all experiments. We evaluate against some of the top algorithms in contrastive learning (SimCLR \cite{simclr}), non-contrastive learning (BYOL \cite{byol}), reconstruction learning (Ti-MAE \cite{timae}), and SSL in ECG (CMSC version of CLOCS \cite{clocs}). We use only this version of CLOCS since the other version requires multiple leads, which is outside the scope of this work. \\

\begin{table}[!ht]
\begin{center}
\caption{Gender Classification Accuracy on PTB-XL}
\begin{tabular}{lcc} 
\toprule
Method & SVC & Random Forest   \\ 
\midrule
BYOL & 0.61 & 0.59  \\
SimCLR & 0.59 & 0.58   \\
Ti-MAE & 0.61 & 0.59  \\
CLOCS & \textbf{0.63} & 0.61  \\
\rowcolor{LightGray}
NERULA & \textbf{0.63} & \textbf{0.64}  \\
\bottomrule
\end{tabular}
\label{tab:gender}
\end{center}
\end{table}

As our work can be seen as an extension across multiple steps from the widely used BYOL \cite{byol} network, we show all additions in our network in Table. \ref{cumulative_table} along with in the incremental performance in F1 scores for each increment. We can view BYOL as a standard non-contrastive learning framework that uses data augmentations like flipping across the x and/or y-axis, time warping, etc. Given below are each of the additions. 
\begin{enumerate}
    \item To this method, we show that using only random masking and generating two masked views of the input - one masked and the other with an inverted mask leads to a 12\% increase in F1 scores from BYOL. 
    \item In this stage, we mask the latent features at every layer to get a masked vector of our final representation. We also replace the encoder which was originally a ResNet \cite{resnet} with a sliding window-based transformer network with masked convolutions \cite{locallead}. This gives us an additional 6\% increase in F1.
    \item Finally, we add our reconstruction network for a dual-pathway architecture that further improves the F1 in arrhythmia classification by 26\%, showing the importance of reconstruction or generative path in representation learning.
\end{enumerate}

We also show the comparison of our network with other state-of-the-art algorithms in Table \ref{comparison_table}. We compare against BYOL \cite{byol}, SimCLR \cite{simclr}, Ti-MAE \cite{timae}, CLOCS \cite{clocs} and also provide the scores of a purely supervised learning network (ResNet \cite{resnet}) to give context. 

\begin{table}[!ht]
\begin{center}
\caption{Age Regression metrics on PTB-XL}
\begin{tabular}{lcccc} 
\toprule
 &  \multicolumn{2}{c}{MLP}  &  \multicolumn{2}{c}{Random Forest}   \\
\cmidrule(lr){2-3} \cmidrule(lr){4-5} 
Method & MAE $\downarrow$ & $R^2$ $\uparrow$ & MAE $\downarrow$ & $R^2$ $\uparrow$ \\ 
\midrule
BYOL & 11.57 & 0.25 & 11.93 & 0.21  \\
SimCLR & 12.54 & 0.12 & 11.80 & 0.22  \\
Ti-MAE & 12.32 & 0.15 & 12.15 & 0.18  \\
CLOCS & 12.20 & 0.16 & 11.79 & 0.22  \\
\rowcolor{LightGray}
NERULA & \textbf{10.46} & \textbf{0.38} & \textbf{10.83} & \textbf{0.34}  \\
\bottomrule
\end{tabular}
\label{tab:age}
\end{center}
\end{table}

In addition to arrhythmia detection, electrocardiogram (ECG) data encapsulate a plethora of vital information pertinent to various other clinical assessments. To explore the versatility and adaptability of our algorithm, we extended our evaluation to include gender classification and age regression tasks utilizing the PTB-XL dataset \cite{ptbxl}. As depicted in Table \ref{tab:gender}, our proposed method demonstrates superior performance in the gender classification tasks compared to existing benchmarks. Furthermore, Table \ref{tab:age} highlights the efficacy of our approach in age regression in both the mean absolute error (MAE) and R-squared ($R^2$) metrics, underscoring the algorithm's robustness and accuracy in predictive modeling across diverse biomedical applications.

In an effort to assess the applicability of our algorithm across various time series analysis tasks, we also evaluate its performance within the domain of Human Activity Recognition (HAR). The dataset utilized for this task is detailed in Section \ref{section:datasets}. Given that our training exclusively involved single-lead ECG data, our approach involved deriving representations for each axis of the accelerometer and gyroscope sensors separately. These representations were then concatenated into a single vector, which subsequently served as the input for the machine-learning model. Although this methodology diverges from conventional practices in handling such data, our objective was to explore the comparative performance of all models when applied to signals distinctly different from ECG data.

Table \ref{tab:har} presents a comprehensive performance evaluation and comparison of our model in the context of activity detection using Inertial Measurement Unit (IMU) sensors. Notably, our model achieved superior scores in this task as well, further evidencing its versatility across a spectrum of time series analysis tasks.

\begin{table}[!ht]
\centering
\caption{Evaluation on Human Activity Recognition (HAR) benchmark with 3-axis accelerometer and gyroscope sensors. All scores are from a Random Forest Classifier fit on top of representations trained from the PhysioNet 2020 dataset.}
\begin{tabular}{lc}
\toprule
Method & Accuracy \\
\midrule
BYOL & 0.85 \\
SimCLR & 0.84 \\
Ti-MAE & 0.67 \\ 
CLOCS & 0.86 \\ 
\rowcolor{LightGray}
NERULA & \textbf{0.87} \\
\bottomrule
\end{tabular}
\label{tab:har}
\end{table}

\section{Conclusion}
In this paper, we presented NERULA, a novel self-supervised learning framework tailored for ECG signal analysis. Our dual-pathway architecture, combining non-contrastive learning and signal reconstruction, represents a novel method in ECG signal processing. The unique 50\% masking strategy, complemented by inverse-masked signals, has demonstrated its effectiveness in enhancing the robustness of the model and dealing with real-world scenarios involving incomplete or corrupted ECG data.

One of the critical aspects of NERULA is its ability to effectively learn from limited and unlabelled data, a common challenge in medical signal processing. Our approach outperforms existing state-of-the-art methods such as BYOL, SimCLR, and CLOCS in downstream linear evaluation tasks, indicating its superiority in extracting meaningful representations from ECG signals. Furthermore, the use of a local attention mechanism in the encoder aligns well with the temporal nature of ECG signals, enabling efficient and focused processing of time-series data.

However, there are several limitations and challenges that we encountered. The choice of a 50\% masking strategy, while empirically effective, was not derived from an extensive hyperparameter tuning process. Additionally, the decision to assign different weights to the losses from the reconstruction and non-contrastive pathways was based on preliminary observations, rather than a systematic exploration. These choices, while pragmatic, highlight the need for further research to optimize these parameters.

\section{Broader Impact}
NERULA’s framework for ECG signal analysis can enhance cardiac health monitoring by providing robust, automated interpretation of ECG data, aiding in early detection of cardiac anomalies. This technology, suitable for wearable devices, could improve patient outcomes and reduce healthcare burden. However, ensuring reliability across diverse populations and maintaining data privacy are crucial for ethical deployment and patient safety.

\newpage
\bibliography{neurips_2024}

\end{document}